\documentclass[aps,showpacs,superscriptaddress,prl,notitlepage,twocolumn,nofootinbibx]{revtex4-1}
\usepackage{amsmath}
\usepackage{amsfonts}
\usepackage{amssymb}
\usepackage[caption=false]{subfig}
\usepackage{appendix}
\usepackage{graphicx}
\usepackage{braket}
\usepackage{xcolor}
\usepackage{hyperref}
\usepackage{floatrow}

\newcommand{\fig}[1]{\mbox{Fig.\ \ref{#1}}}

\begin{document}


  \title{Decoding Holographic Codes with an Integer Optimisation Decoder}
  \author{Robert J. Harris}
  \email{rjh2608@gmail.com}
\affiliation{ARC Centre for Engineered Quantum Systems, School of Mathematics and Physics, The University of Queensland, St Lucia, QLD, 4072, Australia}
\author{Elliot Coupe}
\affiliation{ARC Centre for Engineered Quantum Systems, School of Mathematics and Physics, The University of Queensland, St Lucia, QLD, 4072, Australia}
\author{Nathan A. McMahon}
 \altaddress{Current Address: Friedrich-Alexander-Universität Erlangen-Nürnberg (FAU), Institute of Theoretical Physics, Erlangen, Germany}
\affiliation{ARC Centre for Engineered Quantum Systems, School of Mathematics and Physics, The University of Queensland, St Lucia, QLD, 4072, Australia}
\author{Gavin K. Brennen}
\affiliation{Center for Engineered Quantum Systems, Dept. of Physics and Astronomy, Macquarie University, 2109 New South Wales, Australia}
\author{Thomas M. Stace}
\affiliation{ARC Centre for Engineered Quantum Systems, School of Mathematics and Physics, The University of Queensland, St Lucia, QLD, 4072, Australia}
  \date{\today}
  
\begin{abstract}
We develop a most likely error Pauli error decoding algorithm for stabiliser codes based on general purpose integer optimisation. Using this decoder we analyse the performance of holographic codes against Pauli errors and find numerical evidence for thresholds against Pauli errors for bulk qubits. We compare the performance of  holographic code families of various code rates and find phenomenological Pauli error thresholds ranging from $7\%$ to $16\%$, depending on the code rate.  Additionally we give numerical evidence that specific distance measures of the  codes we consider scales polynomially with number of physical qubits.
\end{abstract} 


  \maketitle
 

Recent interest in the overlap between quantum information and the bulk-boundary correspondence \cite{Harlow2016} has connected holography and error correction \cite{Latorre2015,Pastawski2015,Yoshida2013}. 
In holography, degrees of freedom (DOFs) in a $d+1$ dimensional bulk manifold are encoded in DOFs in a $d$ dimensional boundary manifold leaving substantial redundancy to the encoding. This redundancy makes holography a principle for developing quantum error correcting codes.

\begin{table}[t]
\begin{equation*}
\begin{array}{rlllllllll}
\textrm{index label:} & 1 & 2 & 3 & 4 & 5 & 6 & L & 7 \\ 
 \hline
 & X & X & \mathbb{I} & \mathbb{I} & \mathbb{I} & X & \mathbb{I} & X & \equiv S_1 \\ 
& \mathbb{I} & X & X & X & \mathbb{I} & \mathbb{I} & \mathbb{I} & X& \equiv S_2 \\ 
 & \mathbb{I} & \mathbb{I} & \mathbb{I} & X & X & X & \mathbb{I} & X& \equiv S_3 \\ 
 & Z & Z & \mathbb{I} & \mathbb{I} & \mathbb{I} & Z & \mathbb{I} & Z & \equiv S_4\\ 
 & \mathbb{I} & Z & Z & Z & \mathbb{I} & \mathbb{I} & \mathbb{I} & Z & \equiv S_5\\ 
& \mathbb{I} & \mathbb{I} & \mathbb{I} & Z & Z & Z & \mathbb{I} & Z & \equiv S_6\\
 & X & X & X & X & X & X & X & X & \equiv S_{\bar X}\\ 
& Z & Z & Z & Z & Z & Z & Z & Z & \equiv S_{\bar Z}
\end{array}
\end{equation*}
\caption{The ordered set of \emph{extended} stabilisers and logical operators that define the block-perfect Steane code tensor, $T_{S}^{j_1,...,j_L,j_7}$. The tensor is given by $T_{S}^{j_1,...,j_L,j_7}=\braket{j_1,...,j_L,j_7|T_S}$ where $\ket{T_S}$ is the $+1$ eigenstate of the eight commuting operators, $S_j$, above.}
\label{tab:SteaneTensor}
\end{table}

For practical quantum processing, error correction is essential \cite{Roffe2019,Preskill1998}. 
A good quantum error correcting code (QECC) ideally has four properties: a threshold against physical errors \cite{Barrett2010}; a finite ratio -- the code rate -- of the number of encoded to physical qubits \cite{Roffe2019}, low-weight  parity check operators \cite{Bohdanowicz2019}, and a fault-tolerant path to a universal gate set  \cite{Pastawski2015a}.  In this Letter, we study  code thresholds, rates and distances for a variety of holographic code families, to establish their performance against some of these metrics. 

\citet{Pastawski2015} (HaPPY) developed holographic quantum error correcting codes by tessellating tensor representations of the five qubit QECC \cite{Laflamme1996} on discretised negatively curved space. 
Due to the proportional scaling between bulk and boundary volumes in a hyperbolic tiling, holographic error correcting codes can form finite rate error correcting codes. The rate of the code is tunable via modification of the tiling. HaPPY showed that the code with the maximum achievable rate did not have thresholds against erasures, however a version of the code with reduced rate was shown to have an erasure threshold which is comparable to that of  the surface code \cite{Stace2009, Stace2010, Barrett2010}.

The five qubit QECC was the first code used to generate holographic tilings because it has a \emph{perfect tensor} representation. 
This property is rare amongst known codes. More recently it has been shown that the perfect tensor condition can be relaxed to a  \emph{block perfect} condition \cite{Harris2018}, which coincides with the definition of planar perfect tangles \cite{Berger2018a}. This relaxation allows more choices for the seed codes used to generate the holographic tiling, extending the set of known holographic codes families to include a holographic code based on the Steane  code \cite{Steane1996} (called here the \emph{heptagon code} \cite{Harris2018}).  This was  also shown to have promising erasure thresholds \cite{Harris2018}. 

CSS codes have a transversal  ${\textsc{CNOT}}$ gate; self-dual CSS codes also have transversal Hadamard gates \cite{Preskill1998a}. Additionally it is known how to construct any CSS code with graph states \cite{Bolt2016,Bolt2018a} which is a powerful approach to measurement-based quantum computation. 

Decoding erasures is numerically straightforward \cite{Harris2018}, but optimally decoding computational (e.g.\ Pauli) errors is a numerically hard problem in general \cite{Iyer2015} (though good decoders exist for  specific codes \cite{Sipser1996, Fowler2013,Liu2019}, such as the surface code \cite{Dennis2002,Delfosse2016,Darmawan2018,Delfosse2017} and the colour code \cite{Bombin2018,Kubica2019}). 
Given the relative computational ease of decoding erasure errors, the performance of a new QECC against erasure has been proposed as a performance filter \cite{Delfosse2016,Darmawan2018}. The promising erasure thresholds for holographic codes therefore suggests further study is warranted. 

In this Letter we describe a general purpose decoding algorithm for stabiliser codes based on a global branch and bound integer optimisation \cite{gurobi}. We implement this decoder and numerically compute the performance of various holographic code families, including a new holographic code family based on a small surface code that we introduce here. 
We provide numerical evidence of thresholds against Pauli errors for these holographic code families. Additionally we use related algorithms to find distances of the codes.


\emph{Holographic code construction}: Holographic codes are seeded by an  $[[n,k,d]]$ QECC \cite{Gottesman1997}.  The seed QECC is described by a rank $(n+k)$ tensor, $T$, 
which is the encoding map from logical to physical space $T: \mathcal{H}_{k} \mapsto \mathcal{H}_{n}$. 
To be a valid code, $T$ must be an isometry, $T^\dagger T \propto \mathbb{I}$, which ensures perfect information recovery in the absence of noise.  We define the order-$(n+k)$ seed tensor 
as $T^{j_1,...,j_{n+k}}=\braket{j_1,...,j_{n+k}|T}$, where $\ket{T}$ is the $+1$ eigenstate of  all \emph{extended} stabilisers and logical operators, i.e.\ the extension of the operators to act on an $(n+k)$-qubit Hilbert space.

Conventionally, the tensor indices are arranged to form an encoding map, $T^{j_1,...,j_{n}\leftarrow j_{n+1},...j_{n+k}}:\mathcal{H}_{k} \mapsto \mathcal{H}_{n}$, from the $k$-dimensional logical Hilbert space to the $n$-dimensional physical Hilbert space.  Related codes can be defined with different index partitions, providing the encoding tensor remains an isometry. For example, if $T^{j_2,...,j_{n}\leftarrow j_1,j_{n+1},...j_{n+k}}$ is an isometry, then it is an encoding map for a $[[n-1,k+1,d^\prime]]$ code. 


\emph{Perfect \& Block-Perfect Tensors}: Some classes of tensor that remain an isometry after a permutation of indices. We describe two such classes. Take a tensor with $m$ indices in an ordered set $J=\{j_1,j_2,...,j_{m}\}$. Partition $J$ into ordered subset $A$ and its complement $\overline{A}$, with some permutation $\Pi$ with respect to the reference index, i.e. $\{A|\overline{A}\}=\Pi[J]$. If under all permutations $\Pi$ the tensor remains an isometry then it is known as a \emph{perfect tensor} \cite{Pastawski2015}. 
If the tensor remains an isometry under all cyclic permutations \mbox{$\Pi = \sigma^p$}, where $\sigma^p: j_i \rightarrow j_{i+p}$, we call it \emph{block-perfect}.

\begin{table}[t]
\begin{equation*}
 \begin{array}{rllllllllllllll}
\textrm{index label:}&1& &2& &3& &4& &L& &5\\
 \hline
 &X& &X& &I& &X& &\mathbb{I}& &\mathbb{I}&\equiv S_1& &\\
 &\mathbb{I}& &\mathbb{I}& &X& &X& &\mathbb{I}& &X&\equiv S_2& &\\
 &Z& &\mathbb{I}& &Z& &Z& &\mathbb{I}& &\mathbb{I}&\equiv S_3& &\\
 &\mathbb{I}& &Z& &\mathbb{I}& &Z& &\mathbb{I}& &Z&\equiv S_4& &\\ 
 &X& &\mathbb{I}& &X& &\mathbb{I}& &X& &\mathbb{I}&\equiv S_{\bar{X}}& &\\
 &\mathbb{I}& &\mathbb{I}& &Z& &\mathbb{I}& &Z& &Z&\equiv S_{\bar{Z}}& &
 \end{array}.
 \end{equation*}
\caption{The ordered set of stabilisers and logical operators that define the block-perfect surface code fragment tensor, $T_{\textrm{SCF}}^{j_1,...,j_L,j_5}=\braket{j_1,...,j_L,j_5|T_{\textrm{SCF}}}$ where $\ket{T_{\textrm{SCF}}}$ is the $+1$ eigenstate of $S_j$'s.}
\label{tab:SurfFragStabs}
\end{table}

As an example, the order-8 tensor associated with the Steane code \cite{Steane1996}, $T_{S}^{j_1,...,j_L,j_7}$, is block perfect \cite{Harris2018} with respect to the ordered set of extended operators defined in Table \ref{tab:SteaneTensor}.  The code family based on this seed code will be used in the following numerical results.


%
%

\begin{figure}
\centering
\includegraphics{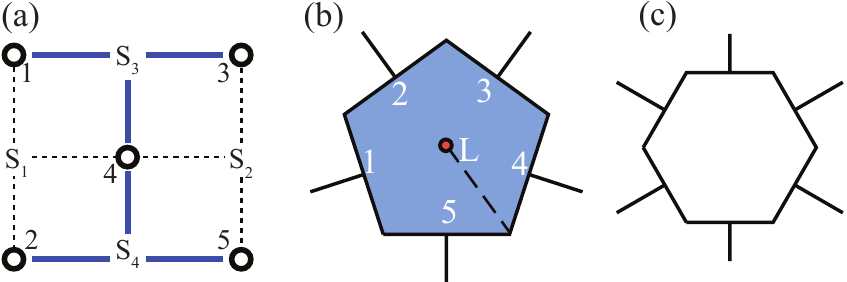}
\caption{(a) Lattice representing the $[[5,1,2]]$ surface code fragment (SCF) including stabilisers. Graphical representations of (b) a $[[5,1]]$ code tensor and (c) the $[[6,0]]$ state associated to a $[[5,1]]$ code. \label{fig:SingleHexagon} \label{fig:SinglePentagon} \label{fig:SurfFragment} }
\end{figure}

We also introduce a new holographic code based on the $[[5,1,2]]$ \emph{surface code fragment} (SCF), defined over 5 qubits, shown in \fig{fig:SurfFragment}a. The  SCF is a block-perfect, error-detecting CSS code. The order-6  SCF tensor, $T_{SCF}^{j_1,...,j_L,j_5}$ defined by the stabilisers in table \ref{tab:SurfFragStabs}, is an isometry for any block partition of indices from any cyclic permutation of $\{ 1,2,3,4,L,5 \}$.



Seed code tensors can be represented graphically as a polygon with a leg for each index of the tensor \cite{Bridgeman2017}. For a $[[n,k,d]]$ error correcting code, the standard graphical representation is an $n$-sided polygon, with a leg on each edge for the physical qubit indices and $k$ \emph{bulk} legs (dots) perpendicular to the face representing each logical qubit. For example \fig{fig:SinglePentagon}b is a representation of a $[[5,1]]$ code.


We conventionally interpret  polygons like \fig{fig:SurfFragment}b as a map from bulk to  planar indices. However, tensor indices can be partitioned into other subsets $A$ and $\overline{A}$: 
if the tensor is an isometry for a particular choice of indices $A$, it describes an encoding map from $A$ to $\overline{A}$. For example, if the tensor in \fig{fig:SinglePentagon}b is block-perfect, then the tensor will be an isometry from  the indices $\{L,5\}$ to $\{1,2,3,4\}$, corresponding to  a $[[4,2]]$ code. 

\begin{figure}
\includegraphics[clip,width=0.48\columnwidth]{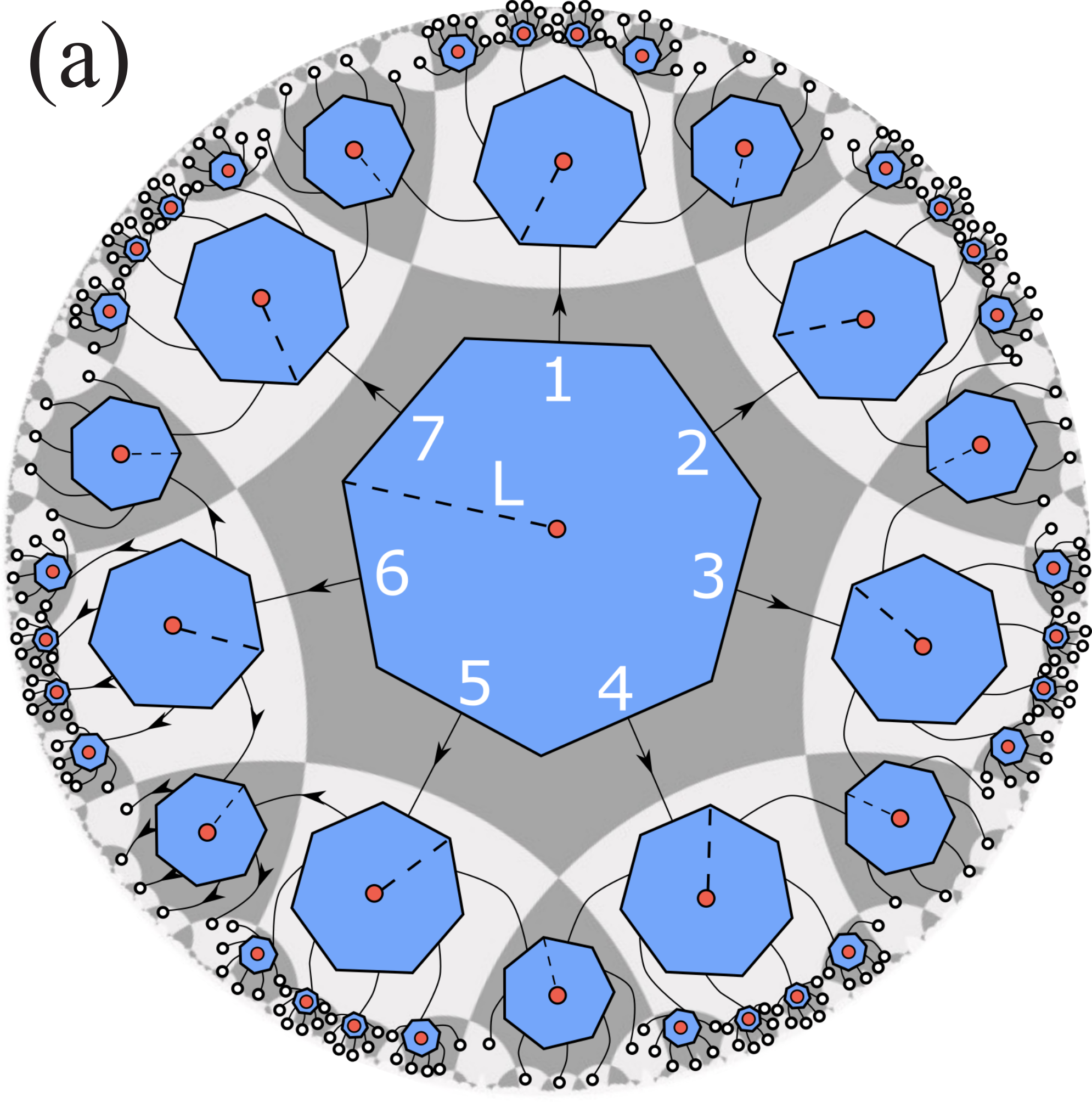}\hfil
\includegraphics[clip,width=0.48\columnwidth]{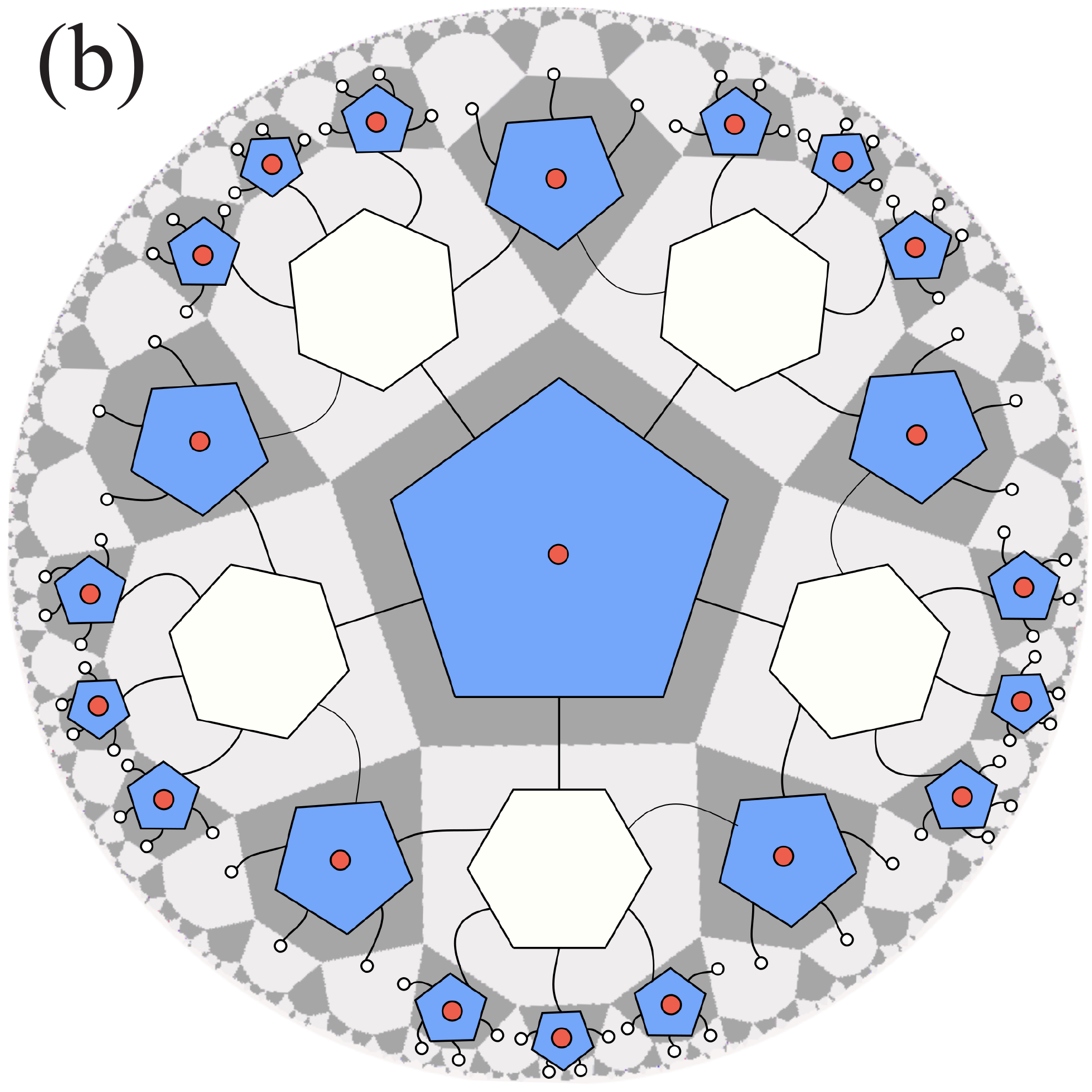}
\caption{Graphical tensor network representation of (a) the Heptagon code \cite{Harris2018} and (b) a reduced-rate Pentagon/Hexagon code. These describe maps from the red logical qubits in the bulk to the white boundary physical qubits. The seed tensor polygons in (b) could represent any $[[5,1]]$ and $[[6,0]]$ seed code tensors such as the SCF or 5-qubit code. }
\label{fig:heptagontiling}\label{fig:PentHex}
\end{figure}

To create holographic codes we tessellate seed codes, represented graphically as polygons-with-legs,  on a discretised 2D hyperbolic space, and contract tensor indices on linked edges. The tessellation is truncated at a particular \emph{radius}, $R$. 
For the heptagon code \cite{Harris2018} we tessellate heptagons which individually represent the Steane code \cite{Steane1996}; the version for $R=3$ is shown in \fig{fig:heptagontiling}a. If every polygon in the tesselation has an associated logical qubit, we call it a \emph{max-rate} holographic code.



Additionally, we consider some tensor networks in which a subset of tiles in the tesselation do not have a logical input, as depicted in \fig{fig:SingleHexagon}c.  Such `blank' tiles do not add bulk logical legs to a network, but allow us to construct \emph{reduced-rate} codes by mixing such tiles amongst others that do have associated bulk indices. 
A particular realisation of this is the pentagon/hexagon code \cite{Pastawski2015}  shown in \fig{fig:PentHex}b, where blank tiles are interspersed in the tesselation.  We note that a given tiling built of order-$(n+k)$ tensors is consistent with any $[[n,k]]$ code.  For instance, the order-6 tensors shown in \fig{fig:PentHex}b  could represent either the $[[5,1,2]]$ surface code fragment from \fig{fig:SurfFragment}a or the $[[5,1,3]]$  five qubit code \cite{Laflamme1996} as introduced by \citet{Pastawski2015}.


Lastly, we will also consider tilings in which only a single seed tensor, at the centre of the tiling, has a logical qubit associated to it, resulting in a  single-logical qubit code with \emph{zero-rate},  \cite{Pastawski2015}.





\emph{Pauli Decoder}: In contrast to erasure decoders, optimally correcting Pauli errors for general stabiliser codes is \#P-complete  \cite{Iyer2015}. 
Here we use general purpose integer optimisation software to perform most likely error (MLE) correction for general stabiliser codes, albeit with a high computational cost. This is useful for evaluating new QECCs, even if it is impractical for real-time experimental implementations. 
The algorithm can be applied to any stabiliser code, with suitable adjustments from the CSS variant described here, which we use to compute numerical thresholds for a Pauli error channel for the heptagon code, the SCF code and the  HaPPY code \cite{Pastawski2015}.

%
We represent operators as binary symplectic vectors (BSV) \cite{Calderbank1997, Gottesman1997,Dehaene2003} for $X$ and $Z$ components defined for $n$-qubit operator $\hat{A}$ as
%
\begin{equation}
\hat{A}={\otimes}_j \hat{X}^{(\underline{a}_X)_j}\hat{Z}^{(\underline{a}_Z)_j}\equiv\hat{X}^{\otimes \underline{a}_X}\hat{Z}^{\otimes \underline{a}_Z}
\end{equation}
where  $\underline{a}_X, \underline{a}_Z \in \mathbb{Z}_2^n$. 
For CSS codes, either $\underline{a}_X$ or $\underline{a}_Z$ will be zero for a given stabiliser or logical operator. Further, for a self-dual CSS code such as the heptagon code 
for any $X$ type stabiliser with $(\underline{a}_X,\underline{a}_Z) = (u,0)$ there is a corresponding $Z$ type stabiliser with $(\underline{a}_X,\underline{a}_Z) = (0,u)$.


To perform decoding we assume
$X$ and $Z$ type errors are I.I.D, similarly to the minimum weight perfect matching decoder for surface codes. However for our simulations we simulate a depolarising error channel, while decoding with the previous assumption. 
For clarity we 
 describe a decoder for a self-dual $[[n,k]]$ CSS QECC subject to $Z$ errors;  $X$ errors are treated in the same way. 
We consider a generic dephasing error given by
\mbox{$
 \hat{\mathcal{E}}=\hat{Z}^{\otimes \underline{\varepsilon}}
$}. 
We define the parity check matrix $\underline{\underline{S}}$, where each stabiliser BSV, $\underline{s}_{j}$, forms a row of the matrix, so that $\underline{\underline{S}}$ is an $\frac{n-k}{2}\times n$ dimensional matrix.  The error syndrome is then given by \mbox{ $\underline{y}=\underline{\underline{S}} \cdot \underline{\varepsilon}$}.
 
 

 
 

From the syndrome we employ an inverse syndrome former (ISF) to find a correction that returns us to the code space. The 
 ISF matrix $\underline{\underline{F}}$ is the pseudoinverse (mod 2) of the parity check matrix, satisfying
\mbox{ $\underline{\underline{F}}^T \cdot \underline{\underline{S}}^T=\mathbb{I}
$}.
That is, each column of the ISF defines an operator that anticommutes with only the corresponding stabiliser. It is not always possible to find an ISF in this way, for example in the toric code each error anticommutes with two stabilisers.
We use $\underline{\underline{F}}$, along with the syndrome to find a \emph{pure error} BSV, $\underline{e}=\underline{\underline{F}} \, \underline{y}$, that satisfies the syndrome, $\underline{y}=\underline{\underline{S}} \cdot \underline{\varepsilon} = \underline{\underline{S}} \cdot \underline{e}$.



  \begin{figure*}[t]
  \includegraphics[clip,width=0.32\columnwidth]{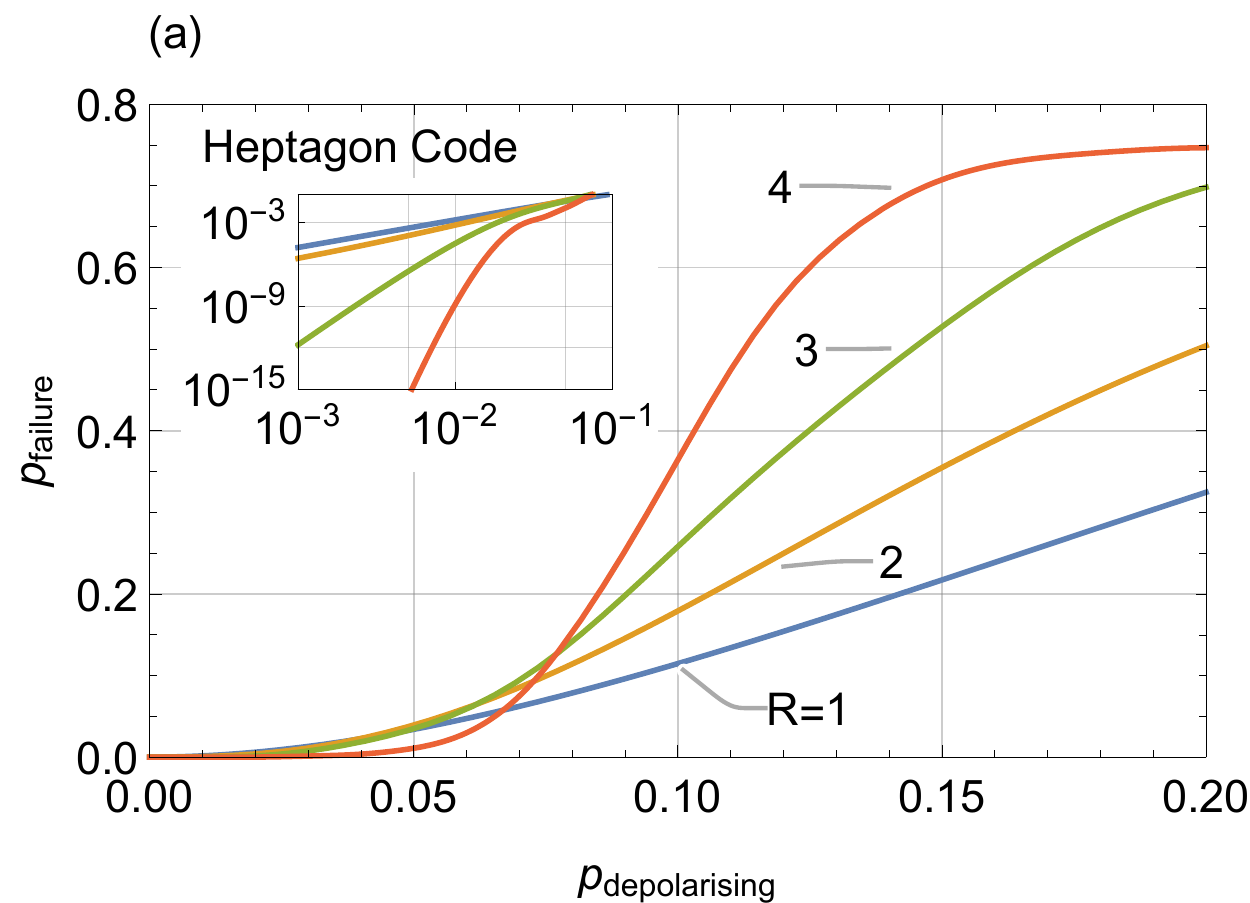}
  \includegraphics[clip,width=0.32\columnwidth]{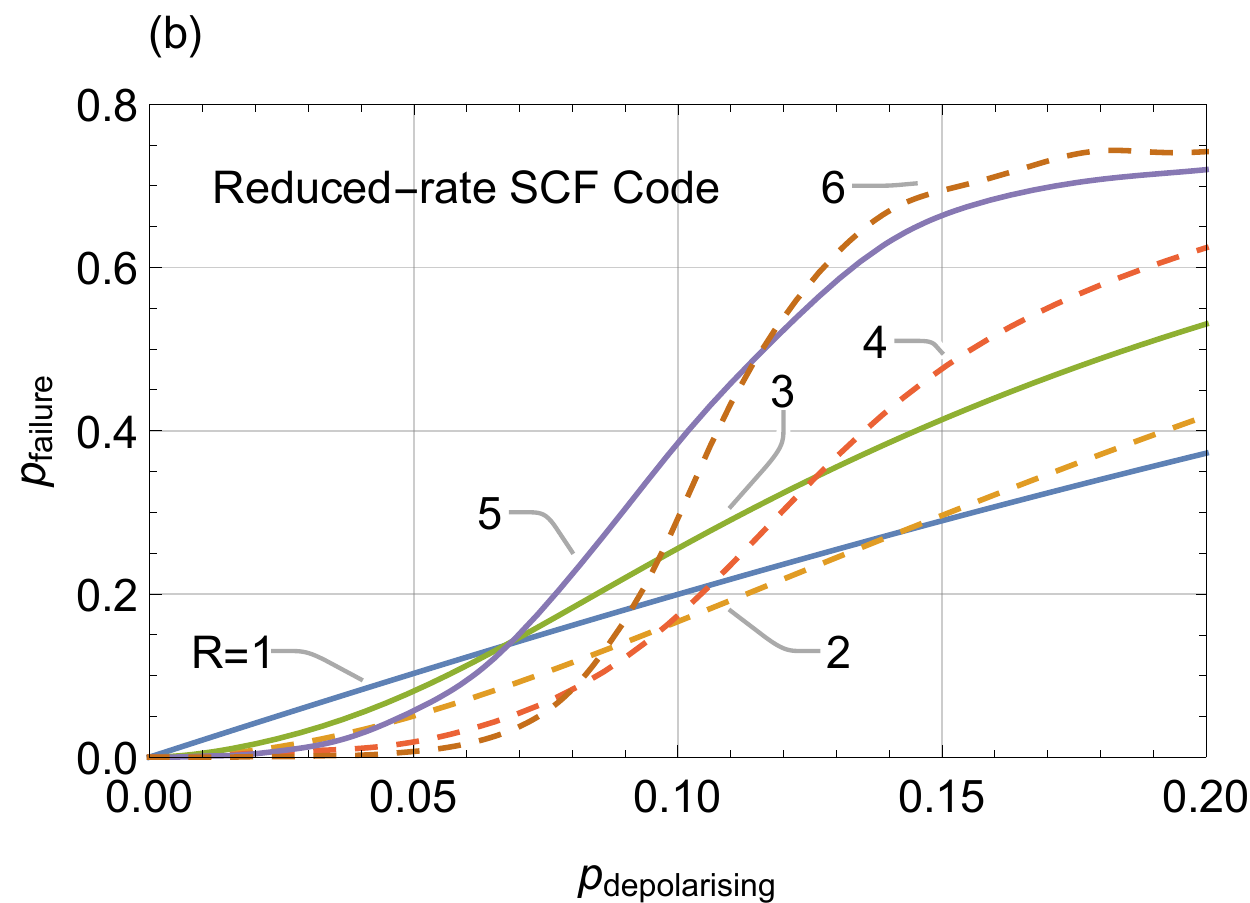}
  \includegraphics[clip,width=0.32\columnwidth]{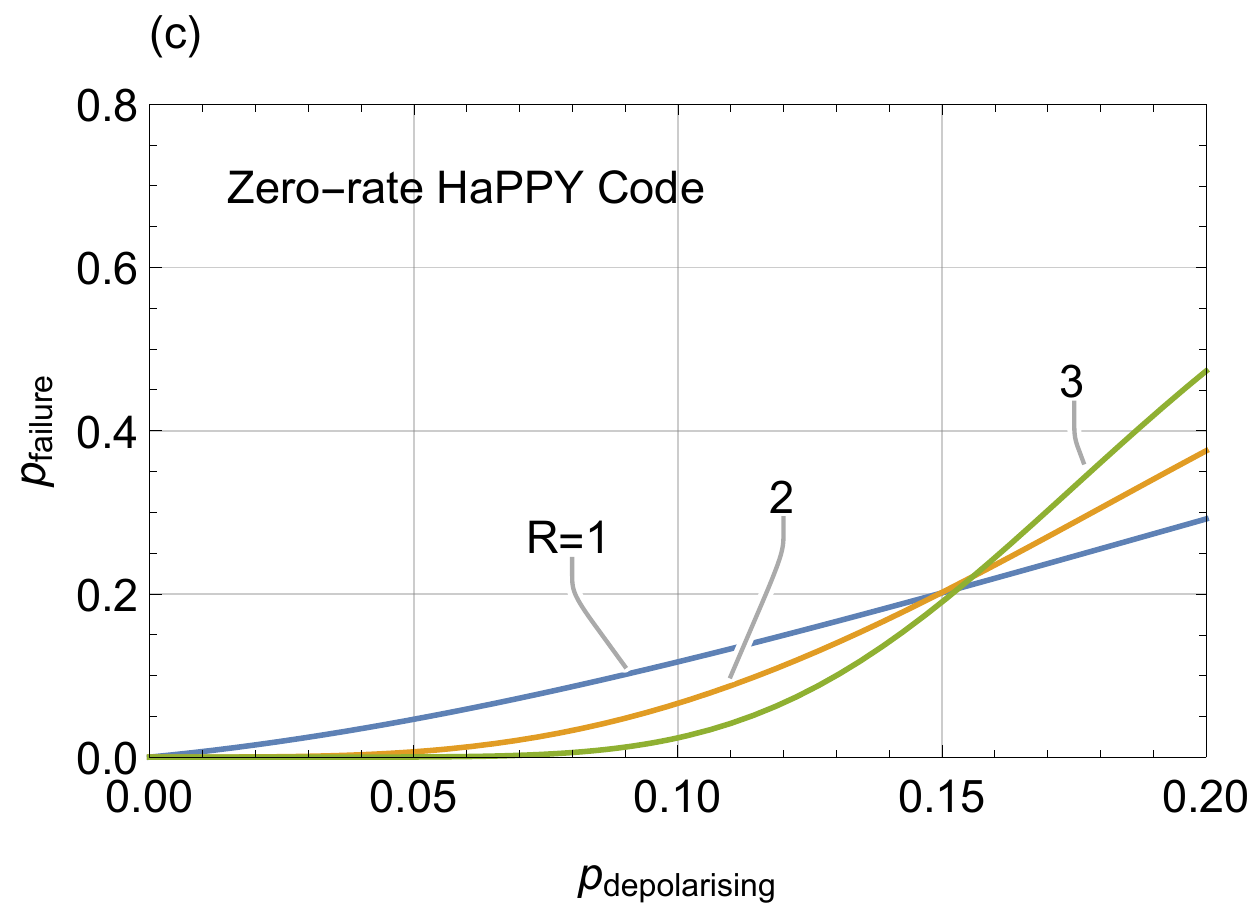}
	
	\caption{Failure probability of the central logical qubit in the a) Max-rate Heptagon Code, b) Reduced-rate SCF Code and c) Zero-rate HaPPY code, in the presence of a depolarising channel.  The inset to fig (a) shows the same data on log-scale to show the performance well-below threshold. Note the error region around the lines are not significantly wider than the lines themselves. The heptagon code raw data with error bars in shown in figure \ref{fig:ErrorGraphs} in the appendix. \label{fig:PauliGraphs} }
\end{figure*}

The complete set of errors that satisfy the syndrome is generated by the product of all combinations of stabilisers and logical operators with $\hat{E}=\hat{Z}^{\otimes \underline{e}}$.  All such errors have BSV of the form 
\begin{equation}
\underline{e}^\prime=\underline{e}+{\sum}_{\ell=1}^{\frac{n-k}{2}}\lambda_\ell \underline{s}_\ell + {\sum}_{m=1}^{k}\mu_m \underline{l}_m,
\label{eq:correction_normal}
\end{equation}
where $\lambda_\ell, \mu_m \in \mathbb{Z}_2$ and  $\underline{l}_m$ is a logical operator BSV. 

 For an IID error model, a MLE decoder  minimises the Hamming weight of $e^\prime$ over $\lambda$ and $\mu$, to find a most likely correction chain
$\underline{c}=\arg\min\limits_{\lambda, \mu \in \mathbb{Z}_2} \textrm{wt}[\underline{e}^\prime]$, 
where $\textrm{wt}$ is the Hamming weight of the vector.  Minimising $\textrm{wt}[\underline{e}^\prime]$ over $\lambda$ and $\mu$  is an integer optimisation problem. 
We implement this optimisation problem using the Gurobi optimisation package \cite{gurobi}. For the examples shown here Gurobi performs global optimisation on the problem.

\emph{Performance Simulations}: We analyse the performance of the  codes against Pauli errors using Monte-Carlo simulation of an IID depolarising error model, with error rate $p$.  
We generate IID patterns of Pauli errors for a fixed number of errors $a=\mathrm{wt}(\underline{\varepsilon})$, from which we compute the syndrome, which we pass to the integer optimising decoder. To decide if the decoder has been successful, we calculate the net error after decoding, which is the product of the original error and the correction, $\underline{\varepsilon}+\underline{c}$. The decoder is successful if the net error acts trivially on the logical code space.

We iterate over all $0\leq a\leq n$ to estimate the recovery probability, $P_{\mathrm{failure}}(a,n)$, and use the binomial formula
\begin{equation}
p_{\textrm{failure}}(p, n)=\sum_a {n\choose a} p^a (1-p)^{n-a} P_{\mathrm{failure}}(a,n),
\label{eq:binomial}
\end{equation}
to calculate the recovery rate for different error rates $p$.

Sampling over many error instances, and different error rates allows us to estimate thresholds in the usual way \cite{Barrett2010}.

We run the decoder for a variety of different tilings and codes.  These include: Heptagon tiling  based on the Steane QECC,  a pentagon tiling  code based on the $[[5,1,3]]$ qubit QECC (i.e.\ the HaPPY code), and a different pentagon tiling based on a $[[5,1,2]]$ surface error-detecting code.  For these broad groups, we encode different numbers of logical qubits, ranging from a single logical qubit at the centre of the tiling which asymptotically has zero rate, to maximum-rate encodings with as many logical qubits as possible.

Further, logical qubits in a given code are not homogeneous: logical qubits encoded near the boundary are protected by fewer stabilisers compared to logical qubits encoded closer to the centre of the tesselation.  As such, here we report threshold figures for the performance of the central logical qubit only.

In Fig \ref{fig:PauliGraphs}a we show the failure probability of the central logical qubit for the max-rate Heptagon code ($r\approx 0.22$) against IID Pauli errors
. With this decoder we see numerical evidence of a threshold near $p\approx7.0\%$ for the central logical qubit.

This threshold is comparable with the surface code which has a threshold of $15.5\%$ for the same noise model with a minimum weight perfect matching decoder \cite{Dennis2002,Wang2010}, albeit with  zero rate. 

We also show the performance of the code  below threshold in the inset of Fig \ref{fig:PauliGraphs}a, which shows that the logical failure rate, $p_{\textrm{failure}}$, decreases exponentially with $p_{\textrm{depolarising}}$.

In Fig \ref{fig:PauliGraphs}b we show the performance of the reduced-rate $[[5,1,2]]$ holographic code. Here we see a pronounced difference between odd radii, where the outer tensors have bulk logical qubits, and even. These are characterised by very different codes rates 
(\mbox{$r_{\textrm{odd}}\approx0.3$} and \mbox{$r_{\textrm{even}}\approx0.09$}), and we find thresholds at $p_\textrm{odd}\approx7.1\%$ and $p_\textrm{even}\approx8.2\%$ respectively.

Additionally we consider the performance of the zero-rate HaPPY code, shown in Fig \ref{fig:PauliGraphs}c. We see evidence of a  threshold here at $16.3\%$. We note that this was the only non-CSS code we analysed with the decoder.  Even though the code has the same tiling as the  SCF (which is CSS),  the decoder run-time and memory consumption was vastly greater than the decoder applied to CSS codes.  As a result, we were only able to estimate the performance of the HaPPY code up to $R=3$, making this threshold estimate merely provisional.


\begin{figure}
\includegraphics[width=0.9\textwidth]{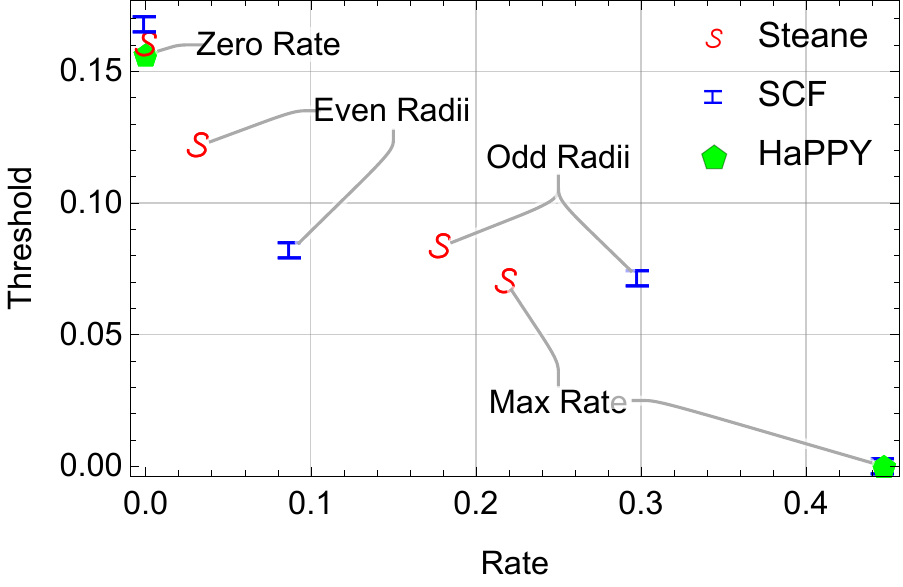}
\caption{A comparison of asymptotic rate and numerically estimated threshold against depolarising errors. of holographic codes seeded with the Heptagon (`Steane') code, the $[[5,1,3]]$  SCF code, and the five qubit $[[5,1,3]]$ code (HaPPY). There is an inverse relationship between threshold and rate.}
\label{fig:ThresholdRatePlot}
\end{figure}



Fig.\ \ref{fig:ThresholdRatePlot} summarises a larger set of numerical simulations, plotting the rate and threshold for a variety of different holographic codes.  We see the expected tradeoff between rate and threshold quite clearly.  Of the codes we have analysed, none dominates the others: increasing rate never increases the threshold, although there are regions where the tradeoff frontier is notably flat.

\emph{Distance}: The decoder also provides a way to find the distance of a code, using the integer optimiser to find the minimum weight logical operator.  
Strictly all maximum rate holographic codes have a fixed distance equal to the distance of the seed code:  this is the distance of the logical qubits closest to the boundary. 
However bulk qubits further from the boundary are better protected than this distance suggests.  A fuller picture is formed by considering the distance of each logical qubit separately.

We define two distance measures  for each  logical qubit. Firstly the logical \emph{bit distance} is the lowest weight logical operator associated to logical qubit $i$ that acts trivially on all other logical qubits, ie.
\begin{equation}
d_i^{(B)}= \min\limits_{\lambda}\textrm{wt}\Big[\underline{l_i}+{\sum}_{j=1}^{J}\lambda_j \underline{s_j}\Big].\nonumber
\end{equation}
Secondly the \emph{word distance} is the minimum weight logical operator that has support on logical qubit $i$, but may act non-trivially on other logical qubits. This distance is equivalent to the distance measure by \citet*{Pastawski2017a}. This is calculated using
\begin{equation}
d_i^{(W)}= \min\limits_{\lambda, \mu} \textrm{wt}\Big[\underline{l_i}+{\sum}_{j=1}^{J}\lambda_j \underline{s_j} + {\sum}_{k\neq i}^{}\mu_k \underline{l_k}\Big].\nonumber
\end{equation}
The word distance counts the shortest weight error that would go undetected but would corrupt the data, while the generally larger bit distance counts how large a logical operator need be to act on an isolated logical qubit, useful e.g. to work out fault tolerant constructions thereof.

\begin{figure}
\includegraphics[width=0.95\textwidth]{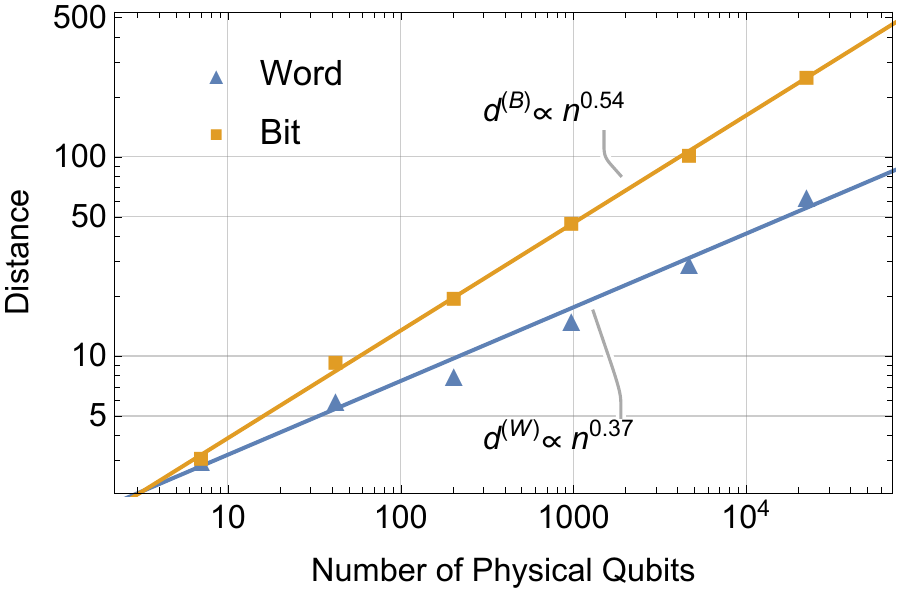}
\caption{The distance of the central logical qubit for the heptagon code at different radii, both the bit and word distance scale exponentially with the number of physical qubits. The fit lines are $d^{(W)}=n^{0.37}$ and $d^{(B)}=n^{0.54}$.\label{fig:HeptagonDistance}}
\end{figure}

Both of these distances can be found using the same integer optimisation package used for decoding.  For different code radii in a given code family we compute the code distance, $d$, and fit a power-law as a function of the number of physical qubits.
 \fig{fig:HeptagonDistance} shows the heptagon code distances (for the central logical qubit) against the number of physical qubits. We see numerical evidence of a power law scaling of distance with the number of physical qubits for both measures.  Similar results are found for other code families.   Table \ref{tab:distacetable} shows the results of numerically found code distances for the other codes studied here, along with the results of power law fits to the numerical results.

We also note that while the numerics here are for the central logical qubit, the important parameter is the bulk qubit distance from the boundary. We say a bulk qubit is positioned at radius $r$, which is the number of tensors between the bulk qubit and the centre. For example, the radius of the central qubit $r=0$ and boundary bulk qubits $r=R-1$. For a given seed tensor and tessellation, the bit distance $d^{(B)}_{R-r}$ is constant for a given value $R-r$, this is also true for word distance.

\begin{table*}
\begin{tabular}[t]{| c || c | c | c || c | c | c || c | c |}
\hline
& \multicolumn{3}{c||}{Heptagon} &  \multicolumn{3}{c||}{Reduced-rate SCF}& \multicolumn{2}{c|}{HaPPY Zero Rate } \\
  \hline
Radius & $n$ & $d^{(B)}$ & $d^{(W)}$ & $n$ & $d^{(B)}$  & $d^{(W)}$ &  $n$ & $d^{(B)}$ \\
\hline
1 & 7 & 3 & 3 & 5 &2 & 2 & 5 & 3\\
\hline
2 & 42 & 9 & 6 & 25 & 4 & 4 & 25 & 9\\
\hline
3 & 203 & 19 & 8 & 75 & 8 & 4 & 95 & 19\\
\hline
4 & 973 & 45 & 15 & 255 & 16 & 8 & 355 & 41\\
\hline
5 & 4662 & 99 & 29 & 745 & 20 & 8 & 1325 & 91\\
\hline
6 & 22337 & 221 & 80 & 2525 & 40 & 16 & 4945 & 321\\
\hline
Asymptotic& $\sim(5+\sqrt{21})^R$ & $\sim n^{0.54\pm0.03}$ & $\sim n^{0.37\pm0.07}$ & $\sim (5+2\sqrt{6})^{R}$ & $\sim n^{0.31\pm0.10}$ & $\sim n^{0.48\pm0.07}$& $\sim(2+\sqrt{3})^R$& $\sim n^{0.65\pm0.08}$ \\
\hline
\end{tabular}
\caption{Numerically optimised bit and word distance of central logical qubits, with the $95\%$ confidence interval, as a function of radius to the boundary qubits. Additionally we show the asymptotic physical qubit count, $n$, as a function of the code radius, $R$, the code rate, $r$, and extrapolated power-law fits (the numerical prefactor is suppressed) to the numerically computed bit, $d^{(B)}$, and word, $d^{(W)}$, distances of the central logical qubit in each code family
.}
\label{tab:distacetable}
\end{table*}





\emph{Conclusions} We have developed a general purpose decoder for stabiliser codes, based on  integer optimisation.  Using this decoder, we have estimated thresholds against IID depolarising noise for different families of holographic codes, and shown that there is a tradeoff between the threshold and the code rate.  The moderately high thresholds against phenomenological depolarising noise are comparable to the surface code, together with finite rates,  may offer an alternative  for building quantum processors, however the development of an efficient decoder that also has thresholds is essential for implementation. The holographic code stabilisers  are non-local, so present challenges in implementation, however generating these codes with low-valence, local cluster states is the subject of ongoing work. 

\begin{acknowledgements}
We thank Stephen Wright, Fred Roosta-Khorasani and David Poulin for useful discussions. 
This work was supported by the Australian Research Council Centre of Excellence for Engineered Quantum Systems (Grant No. CE 170100009), and the Asian Office of Aerospace Research and Development (AOARD) grant FA2386-18-1-4027.
\end{acknowledgements}

  \bibliography{ErrorCorrection,misc}

  \appendix
  \section{Appendix}
  
\begin{figure*}[t]
  \includegraphics[clip,width=0.32\columnwidth]{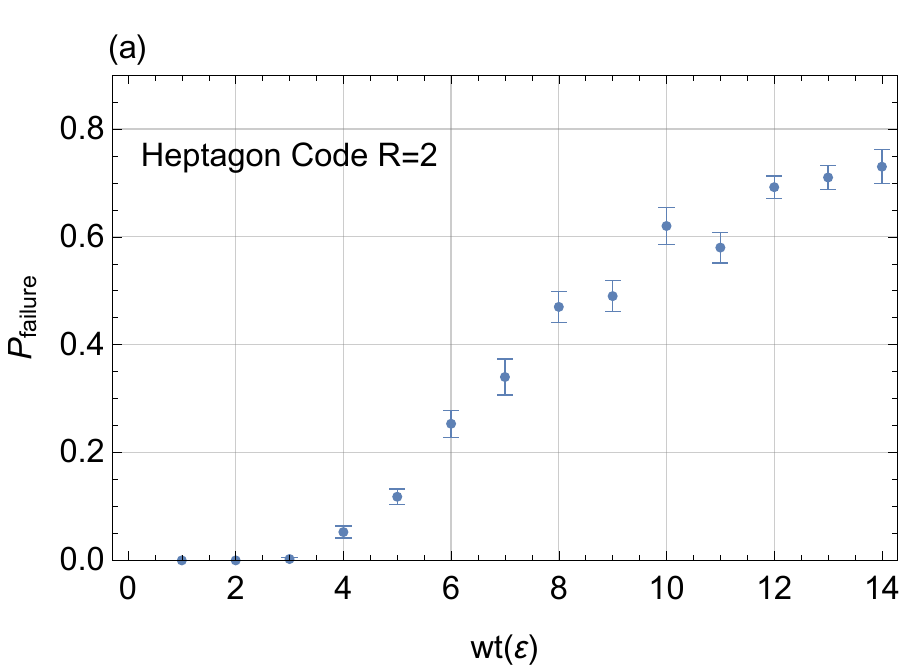}
  \includegraphics[clip,width=0.32\columnwidth]{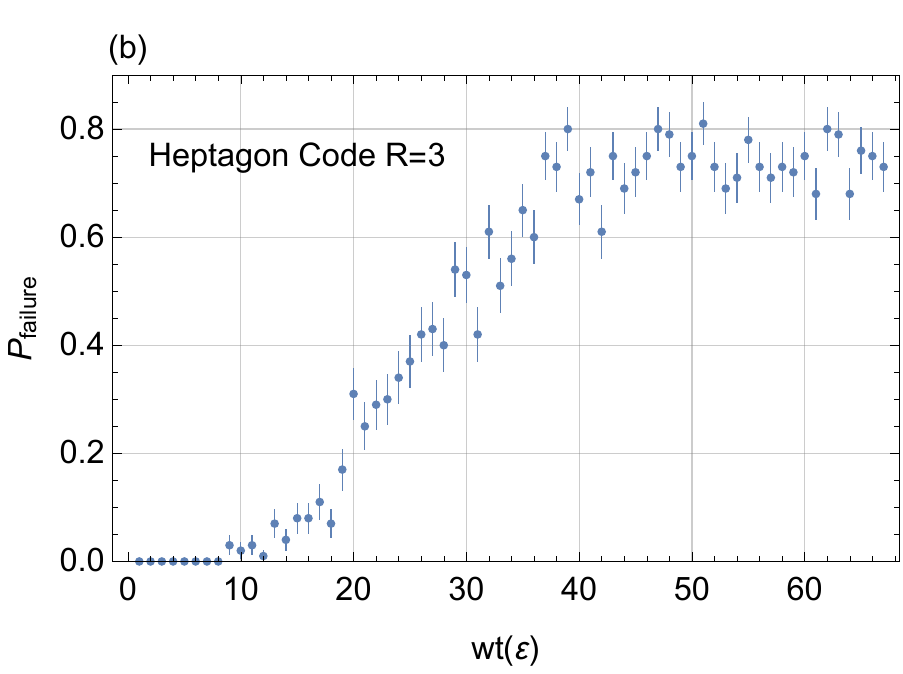}
  \includegraphics[clip,width=0.32\columnwidth]{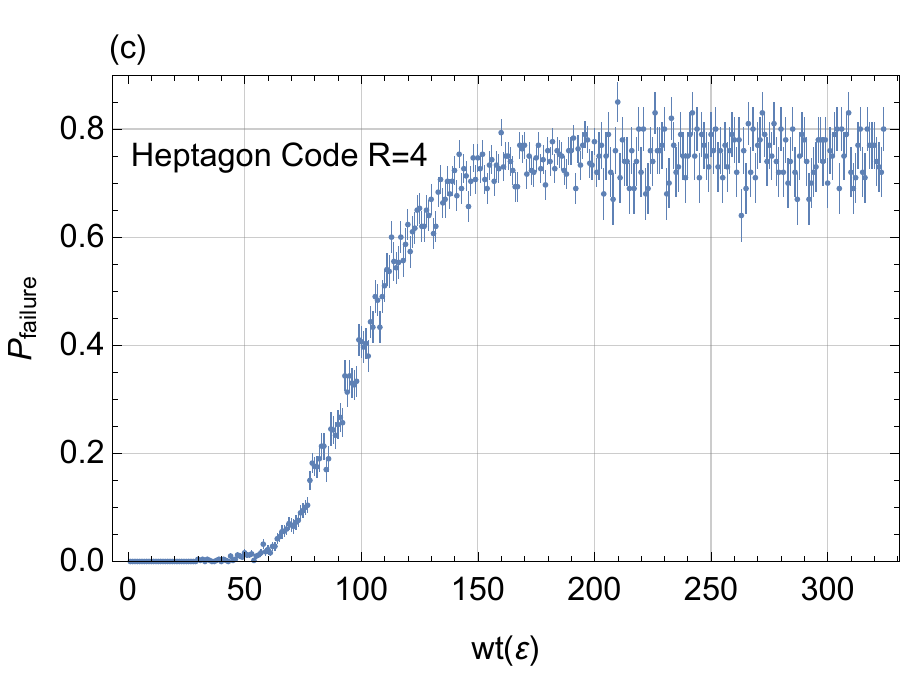}
	
	\caption{Failure probability of the central logical qubit, with error bars, in the max-rate Heptagon code for radius a) $R=2$, b) $R=3$ and c) $R=4$, in the presence of a depolarising channel of fixed weight $\textrm{wt}(\varepsilon)$. 
	\label{fig:ErrorGraphs} }
\end{figure*}

  We now briefly mention the uncertainties in our analysis. For the failure probability for fixed weight $P_{\textrm{failure}}$, we estimate the uncertainty in the results as
  \begin{equation}
      \sigma = \sqrt{\frac{P_{\textrm{failure}}(1-P_{\textrm{failure}})}{m}},
      \label{eq:error}
  \end{equation}
  where $m$ is the number of samples. In figure \ref{fig:ErrorGraphs} we show the estimate $P_{\textrm{failure}}$ for the max-rate heptagon code with radii $2-4$, along with error bars calulated using equation \ref{eq:error}.
  
\end{document}